\documentclass[
aps,pra,
reprint,
a4paper,
superscriptaddress,
longbibliography,
preprintnumbers,
]{revtex4-1}
\usepackage[utf8]{inputenc}
\usepackage[T1]{fontenc}

\usepackage{amsmath,amsthm,amsfonts}
\usepackage{braket}
\usepackage{graphicx}
\usepackage{hyperref}

\hypersetup{colorlinks=true}
\hypersetup{citecolor=blue}
\hypersetup{urlcolor=blue}

\DeclareMathOperator{\diag}{diag}
\newcommand{\bisc}[2]{{[{#1}\text{\bf~:~}{#2}]}}
\newcommand{\omitted}{{\rule{1ex}{.4pt}}}
\newcommand{\proj}[1]{{\ket{#1}\!\!\bra{#1}}}
\newcommand{\reals}{{\mathbb R}}

\begin{document}


\title{Proposed experiment to test fundamentally binary theories}


\author{Matthias~Kleinmann}
\email{matthias_kleinmann001@ehu.eus}
\affiliation{Department of Theoretical Physics, University of the Basque 
Country UPV/EHU, P.O.~Box 644, E-48080 Bilbao, Spain}

\author{Tamás~Vértesi}
\email{tvertesi@atomki.mta.hu}
\affiliation{Institute for Nuclear Research, Hungarian Academy of Sciences, 
H-4001 Debrecen, P.O.~Box 51, Hungary}

\author{Adán~Cabello}
\email{adan@us.es}
\affiliation{Departamento de Física Aplicada II, Universidad de Sevilla, 
E-41012 Sevilla, Spain}


\begin{abstract}
Fundamentally binary theories are nonsignaling theories in which measurements 
 of many outcomes are constructed by selecting from binary measurements.
They constitute a sensible alternative to quantum theory and have never been 
 directly falsified by any experiment.
Here we show that fundamentally binary theories are experimentally testable 
 with current technology.
For that, we identify a feasible Bell-type experiment on pairs of entangled 
 qutrits.
In addition, we prove that, for any $n$, quantum $n$-ary correlations are not 
 fundamentally $(n-1)$-ary.
For that, we introduce a family of inequalities that hold for fundamentally 
 $(n-1)$-ary theories but are violated by quantum $n$-ary correlations.
\end{abstract}

\maketitle


\section{Introduction}


Quantum theory (QT) is the most successful theory physicists have ever devised.
Still, there is no agreement on which physical reasons force its formalism 
 \cite{FS16}.
It is therefore important to test ``close-to-quantum'' alternatives, defined as 
 those which are similar to QT in the sense that they have entangled states, 
 incompatible measurements, violation of Bell inequalities, and no experiment 
 has falsified them, and sensible in the sense that they are in some aspects 
 simpler than QT.
Examples of these alternatives are theories allowing for almost quantum 
 correlations \cite{NGHA15}, theories in which measurements are fundamentally 
 binary \cite{KC16}, and theories allowing for a higher degree of 
 incompatibility between binary measurements \cite{BHSS13}.

Each of these alternatives identifies a particular feature of QT that we do not 
 fully understand and, as a matter of fact, may or may not be satisfied by 
 nature.
For example, we still do not know which principle singles out the set of 
 correlations in QT \cite{Cabello15}.
In contrast, the set of almost quantum correlations satisfies a list of 
 reasonable principles and is simple to characterize \cite{NGHA15}.
Similarly, we do not know why in QT there are measurements that cannot be 
 constructed by selecting from binary measurements \cite{KC16}.
However, constructing the set of measurements of the theory would be simpler if 
 this would not be the case.
Finally, we do not know why the degree of incompatibility of binary 
 measurements in QT is bounded as it is, while there are theories that are not 
 submitted to such a limitation \cite{BHSS13}.

Unfortunately, we do not yet have satisfactory answers to these questions.
Therefore, it is important to test whether nature behaves as predicted by QT 
 also in these particular aspects.
However, this is not an easy task.
Testing almost quantum theories is difficult because we still do not have a 
 well-defined theory; thus, there is not a clear indication on how we should 
 aim our experiments.
Another reason, shared by theories with larger binary incompatibility, is that 
 the only way to test them is by proving that QT is wrong, which is, arguably, 
 very unlikely.
The case of fundamentally binary theories is different.
We have explicit theories \cite{KC16} and we know that fundamentally binary 
 theories predict supraquantum correlations for some experiments but subquantum 
 correlations for others.
That is, if QT is correct, there are experiments that can falsify fundamentally 
 binary theories \cite{KC16}.
The problem is that all known cases of subquantum correlations require 
 visibilities that escape the scope of current experiments.

This is particularly unfortunate now that, after years of efforts, we have 
 loophole-free Bell inequality tests \cite{HBD15,GVW15,SMC15,HKB16,W16}, tests 
 touching the limits of QT \cite{PJC15,CLBGK15}, and increasingly sophisticated 
 experiments using high-dimensional two-photon entanglement 
 \cite{VWZ02,GJVWZ06,DLBPA11}.
Therefore, a fundamental challenge is to identify a feasible experiment 
 questioning QT beyond the local realistic theories \cite{Bell64}.

The main aim of this work is to present a feasible experiment capable of 
 excluding fundamentally binary theories.
In addition, the techniques employed to identify that singular experiment will 
 allow us to answer a question raised in Ref.~\cite{KC16}, namely, whether or 
 not, for some $n$, quantum $n$-ary correlations are fundamentally $(n-1)$-ary.


\subsection{Device-independent scenario}


Consider a bipartite scenario where two observers, Alice and Bob, perform 
 independent measurements on a joint physical system.
For a fixed choice of measurements $x$ for Alice and $y$ for Bob, $P(a,b|x,y)$ 
 denotes the joint probability of Alice obtaining outcome $a$ and Bob obtaining 
 outcome $b$.
We assume that both parties act independently in the sense that the marginal 
 probability for Alice to obtain outcome $a$ does not depend on the choice of 
 Bob's measurement $y$, i.e., $\sum_b P(a,b|x,y)\equiv 
 P(a,\omitted|x,\omitted)$, and analogously $\sum_a P(a,b|x,y)\equiv 
 P(\omitted,b|\omitted,y)$.
These are the nonsignaling conditions, which are obeyed by QT whenever both 
 observers act independently, in particular, if the operations of the observers 
 are spacelike separated.
However, QT does not exhaust all possible correlations subject to these 
 constraints \cite{PR94}.

The strength of this scenario lies in the fact that the correlations can be 
 obtained without taking into account the details of the experimental 
 implementation and hence it is possible to make statements that are 
 independent of the devices used.
This device-independence allows us to test nature without assuming a particular 
 theory---such as QT---for describing any of the properties of the measurement 
 setup.
This way, it is also possible to make theory-independent statements and, in 
 particular, to analyze the structure of any probabilistic theory that obeys 
 the nonsignaling conditions.


\subsection{Fundamentally binary theories}


One key element of the structure of any probabilistic theory was identified in 
 Ref.~\cite{KC16} and concerns how the set of measurements is constructed, 
 depending on the number of outcomes.
According to Ref.~\cite{KC16}, it is plausible to assume that a theory 
 describing nature has, on a fundamental level, only measurements with two 
 outcomes while situations where a measurement has more outcomes are achieved 
 by classical postprocessing of one or several two-outcome measurements.
To make this a consistent construction, it is also admissible that the 
 classical postprocessing depends on additional classical information and, in 
 the bipartite scenario, this classical information might be correlated between 
 both parties.
The total correlation attainable in such a scenario are the binary nonsignaling 
 correlations, which are characterized by the convex hull of all nonsignaling 
 correlations obeying $P(a,\omitted|x,\omitted)= 0$ for all measurements $x$ 
 and all but two outcomes $a$, and $P(\omitted,b|\omitted,y) = 0$ for all 
 measurements $y$ and all but two outcomes $b$.
The generalization to $n$-ary nonsignaling correlations is straightforward.

In Ref.~\cite{KC16}, it was shown that for no $n$ the set of $n$-ary nonlocal 
 correlations covers all the set of quantum correlations.
Albeit this being a general result, the proof in Ref.~\cite{KC16} has two 
 drawbacks:
(i) It does not provide a test which is experimentally feasible.
(ii) It does not allow us to answer whether or not quantum $n$-ary correlations 
 are still fundamentally $(n-1)$-ary.
For example, the proof in Ref.~\cite{KC16} requires {10}-outcome quantum 
 measurements for excluding the binary case.
In this work, we address both problems and provide
(i') an inequality that holds for all binary nonsignaling correlations, but can 
 be violated using three-level quantum systems (qutrits) with current 
 technology, and
(ii') a family of inequalities obeyed by $(n-1)$-ary nonsignaling correlations 
 but violated by quantum measurements with $n$ outcomes.


\section{Results}


\subsection{Feasible experiment to test fundamentally binary theories}


We first consider the case where Alice and Bob both can choose between two 
 measurements, $x=0,1$ and $y=0,1$, and each measurement has three outcomes 
 $a,b=0,1,2$.
For a set of correlations $P(a,b|x,y)$, we define
\begin{equation}
 I_a=\sum_{k,x,y=0,1} (-1)^{k+x+y}P(k,k|x,y),
\end{equation}
 where the outcomes with $k=2$ do not explicitly appear.
With the methods explained in Sec.~\ref{polymeth}, we find that, up to 
 relabeling of the outcomes,
\begin{equation}\label{ineqa}
 I_a\le 1
\end{equation}
 holds for nonsignaling correlations if and only if the correlations are 
 fundamentally binary.
However, according to QT, the inequality in Eq.~\eqref{ineqa} is violated, and 
 a value of
\begin{equation}\label{qvaluea}
 I_a= 2(2/3)^{3/2}\approx 1.0887
\end{equation}
 can be achieved by preparing a two-qutrit system in the pure state
\begin{equation}
 \ket\psi=\frac{1}{2}(\sqrt{2}\ket{00}+ \ket{11}-\ket{22})
\end{equation}
 and choosing the measurements $x,y=0$ as $M_{k|0}= V\proj{k}V^\dag$, and the 
 measurements $x,y=1$ as $M_{k|1}= U\proj{k}U^\dag$, where, in canonical matrix 
 representation,
\begin{equation}
 V=\frac1{\sqrt{12}}\begin{pmatrix} 2 & 2 & 2 \\
 -\sqrt{3}-1 & \sqrt{3}-1 & 2 \\
 \sqrt{3}-1 & -\sqrt{3}-1 & 2
\end{pmatrix},
\end{equation}
 and $U=\diag(-1,1,1)V$.

Using the second level of the Navascués--Pironio--Acín (NPA) hierarchy 
 \cite{NPA07}, we verify that the value in Eq.~\eqref{qvaluea} is optimal 
 within our numerical precision of $10^{-6}$.
The visibility required to observe a violation of the inequality in 
 Eq.~\eqref{ineqa} is $91.7\%$, since the value for the maximally mixed state 
 is $I_a=0$.
The visibility is defined as the minimal $p$ required to obtain a violation 
 assuming that the prepared state is a mixture of the target state and a 
 completely mixed state, $\rho_{\rm prepared} = p \proj\psi + (1-p) \rho_{\rm 
 mixed}$.

We show in Sec.~\ref{polymeth} that the inequality in Eq.~\eqref{ineqa} holds 
 already if only one of the measurements of either Alice or Bob is 
 fundamentally binary.
Therefore, the violation of the inequality in Eq.~\eqref{ineqa} allows us to 
 make an even stronger statement, namely, that none of the measurements used is 
 fundamentally binary, thus providing a device-independent certificate of the 
 genuinely ternary character of all measurements in the experimental setup.

The conclusion at this point is that the violation of the inequality in 
 Eq.~\eqref{ineqa} predicted by QT could be experimentally observable even 
 achieving visibilities that have been already attained in previous 
 Bell-inequality experiments on qutrit--qutrit systems 
 \cite{VWZ02,GJVWZ06,DLBPA11}.
It is important to point out that, in addition, a compelling experiment 
 requires that the local measurements are implemented as measurements with 
 three outcomes rather than measurements that are effectively two-outcome 
 measurements.
That is, there should be a detector in each of the three possible outcomes of 
 each party.
The beauty of the inequality in Eq.~\eqref{ineqa} and the simplicity of the 
 required state and measurements suggest that this experiment could be carried 
 out in the near future.


\subsection{Quantum $n$-ary correlations are not fundamentally $(n-1)$-ary}


If our purpose is to test whether or not one particular measurement is 
 fundamentally binary (rather than all of them), then it is enough to consider 
 a simpler scenario where Alice has a two-outcome measurement $x=0$ and a 
 three-outcome measurement $x=1$, while Bob has three two-outcome measurements 
 $y=0,1,2$.
We show in Sec.~\ref{polymeth} that for the combination of correlations
\begin{equation}\label{ieb}
 I_b=-P(0,\omitted|0,\omitted)+\sum_{k=0,1,2}[P(0,0|0,k)-P(k,0|1,k)],
\end{equation}
 up to relabeling of the outcomes and Bob's measurement settings,
\begin{equation}\label{ineqb}
 I_b\le 1
\end{equation}
 holds for nonsignaling correlations if and only if the correlations are 
 fundamentally binary.
According to QT, this bound can be violated with a value of
\begin{equation}\label{qvalueb}
 I_b=\sqrt{16/15}\approx 1.0328,
\end{equation}
 by preparing the state
\begin{equation}
 \ket\psi=\frac1{\sqrt{(3\zeta+1)^2+2}}(\ket{00}+\ket{11}+\ket{22}+ 
\zeta\ket\phi\!\ket\phi),
\end{equation}
 where $\zeta= -\frac13+\frac16\sqrt{10\sqrt{15}-38}\approx -0.19095$, 
 $\ket\phi=\ket0+\ket1+\ket2$, and choosing Alice's measurement $x=0$ as 
 $A_{0|0}=\openone-A_{1|0}$, $A_{1|0}=\proj{\phi}/3$, and measurement $x=1$ as 
 $A_{k|1}=\proj k$, for $k=0,1,2$, and Bob's measurements $y=0,1,2$ as 
 $B_{0|y}=\openone-B_{1|y}$ and $B_{1|k}=\proj{\eta_k}/\braket{\eta_k|\eta_k}$, 
 where $\ket{\eta_k}=\ket{k}+\xi\ket\phi$, for $k=0,1,2$, and $\xi = 
 -\frac13+\frac16\sqrt{6\sqrt{15}+22}\approx 0.78765$.
[Another optimal solution is obtained by flipping the sign before the 
 $(\frac16\sqrt{\,})$-terms in $\xi$ and $\zeta$, yielding $\xi\approx -1.4543$ 
 and $\zeta\approx -0.47572$.]

We use the third level of the NPA hierarchy to confirm that, within our 
 numerical precision of $10^{-6}$, the value in Eq.~\eqref{qvalueb} is optimal.
Notice, however, that the visibility required to observe a violation of the 
 inequality in Eq.~\eqref{ineqb} is $96.9\%$.
This contrasts with the $91.7\%$ required for the inequality in 
 Eq.~\eqref{ineqa} and shows how a larger number of outcomes allows us to 
 certify more properties with a smaller visibility.

Nevertheless, what is interesting about the inequality in Eq.~\eqref{ineqb} is 
 that it is a member of a family of inequalities and this family allows us to 
 prove that, for any $n$, quantum $n$-ary correlations are not fundamentally 
 $(n-1)$-ary, a problem left open in Ref.~\cite{KC16}.
For that, we modify the scenario used for the inequality in Eq.~\eqref{ineqb}, 
 so that now Alice's measurement $x=1$ has $n$ outcomes, while Bob has $n$ 
 measurements with two outcomes.
We let $I_b^{(n)}$ be as $I_b$ defined in Eq.~\eqref{ieb}, with the only 
 modification that in the sum, $k$ takes values from $0$ to $n-1$.
Then,
\begin{equation}\label{ineqc}
 I_b^{(n)}\le n-2
\end{equation}
 is satisfied for all fundamentally $(n-1)$-ary correlations.
The proof is given in Sec.~\ref{proof}.
Clearly, the value $I_b^{(n)}=n-2$ can already be reached by choosing the fixed 
 local assignments where all measurements of Alice and Bob always have outcome 
 $a,b=0$.
According to QT, it is possible to reach values of $I_b^{(n)}> (n-2)+1/(4n^3)$, 
 as can be found by generalizing the quantum construction from above to 
 $n$-dimensional quantum systems with $\xi=\sqrt2$ and $\zeta= 
 -1/n+1/(\sqrt2n^2)$.
Thus, the $(n-1)$-ary bound is violated already by $n$-ary quantum 
 correlations.
Note, that the maximal quantum violation is already very small for $n=4$ as the 
 bound from the third level of the NPA hierarchy is $I_b^{(4)}<2.00959$.


\section{Methods}


\subsection{Restricted nonsignaling polytopes}\label{polymeth}


We now detail the systematic method that allows us to obtain the inequalities 
 in Eqs.~\eqref{ineqa}, \eqref{ineqb}, and \eqref{ineqc}.
We write $S=\bisc{a_1, a_2,\dotsc, a_n}{b_1, b_2,\dotsc, b_m}$ for the case 
 where Alice has $n$ measurements and the first measurement has $a_1$ outcomes, 
 the second $a_2$ outcomes, etc., and similarly for Bob and his $m$ 
 measurements with $b_1$, $b_2$,\dots, outcomes.
The nonsignaling correlations for such a scenario form a polytope $C(S)$.
For another bipartite scenario $S'$ we consider all correlations $P'\in C(S')$ 
 that can be obtained by local classical postprocessing from any $P\in C(S)$.
The convex hull of these correlations is again a polytope and is denoted by 
 $C(S\rightarrow S')$.

The simplest nontrivial polytope of fundamentally binary correlations is then 
 $C(\bisc{2,2}{2,2}\rightarrow \bisc{3,3}{3,3})$.
We construct the vertices of this polytope and compute the {468} facet 
 inequalities (i.e., tight inequalities for fundamentally binary correlations) 
 with the help of the Fourier-Motzkin elimination implemented in the software 
 \texttt{porta} \cite{porta}.
We confirm the results by using the independent software \texttt{ppl} 
 \cite{ppl}.
Up to relabeling of the outcomes, only the facet $I_a\le 1$ is not a face of 
 the set the nonsignaling correlations $C(\bisc{3,3}{3,3})$, which concludes 
 our construction of $I_a$.
In addition, we find that
\begin{equation}\label{coneq}
C(\bisc{2,3}{3,3})= C(\bisc{2,2}{2,2}\rightarrow \bisc{2,3}{3,3}),
\end{equation}
 and therefore the inequality in Eq.~\eqref{ineqa} holds for all nonsignaling 
 correlations where at least one of the measurements is fundamentally binary.

As a complementary question we consider the case where only a single 
 measurement has three outcomes.
According to Eq.~\eqref{coneq}, the smallest scenarios where such a 
 verification is possible are $\bisc{2,3}{2,2,2}$ and $\bisc{2,2}{2,2,3}$.
We first find that $C(\bisc{2,2}{3,3,3})= C(\bisc{2,2}{2,2,2}\rightarrow 
 \bisc{2,2}{3,3,3})$, i.e., even if all of Bob's measurements would be 
 fundamentally ternary, the correlations are always within the set of 
 fundamentally binary correlations.
Hence, we investigate the polytope $C(\bisc{2,2}{2,2,2}\rightarrow 
 \bisc{2,3}{2,2,2})$ and its {126} facets.
Up to symmetries, only the facet $I_b\le 1$ is not a face of 
 $C(\bisc{2,3}{2,2,2})$.

Our method also covers other scenarios.
As an example we study the polytope $C(\bisc{2,4}{2,4}\rightarrow 
 \bisc{2,2,2}{2,2,2})$ with its {14052} facets.
In this case, the four-outcome measurements have to be distributed to 
 two-outcome measurements (or the two-outcome measurement is used twice).
Hence, this scenario is equivalent to the requirement that for each party at 
 least two of the three measurements are compatible.
The polytope has, up to relabeling, {10} facets that are not a face of 
 $C(\bisc{2,2,2}{2,2,2})$.
According to the fourth level of the NPA hierarchy, two of the facets may 
 intersect with the quantum correlations.
While for one of them the required visibility (with respect to correlations 
 where all outcomes are equally probable) is at least $99.94\%$, the other 
 requires a visibility of at least $97.88\%$.
This latter facet is $I_c\le 0$, where
\begin{multline}
I_c=-P(10|00)-P(00|01)-P(00|10)-P(00|11)\\
 -P(10|12)-P(01|20)-P(01|21)+P(00|22).
\end{multline}
For arbitrary nonsignaling correlations, $I_c\le 1/2$ is tight, while within 
 QT, $I_c< 0.0324$ must hold.
We can construct a numeric solution for two qutrits which matches the bound 
 from the third level of the NPA hierarchy up to our numerical precision of 
 $10^{-6}$.
The required quantum visibility then computes to $97.2\%$.
The quantum optimum is reached for measurements $A_{0|k}=\proj{\alpha_k}$, 
 $A_{1|k}=\openone -A_{0|k}$, and $B_{0|k}=\proj{\beta_k}$, $B_{1|k}=\openone 
 -B_{0|k}$, where all $\ket{\alpha_k}$ and $\ket{\beta_k}$ are normalized and 
 $\braket{\alpha_0|\alpha_1}\approx 0.098$, $\braket{\alpha_0|\alpha_2}\approx 
 0.630$, $\braket{\alpha_1|\alpha_2}\approx 0.572$, and 
 $\braket{\beta_k|\beta_\ell}\approx 0.771$ for $k\ne \ell$.
A state achieving the maximal quantum value is $\ket\psi\approx 
 0.67931\ket{00}+0.67605\ket{11}+0.28548\ket{22}$.
Note, that $I_c\approx 0.0318$ can still be reached according to QT, when Alice 
 has only two incompatible measurements by choosing 
 $\braket{\alpha_0|\alpha_1}= 0$.
Curiously, the facet $I_c\le 0$ is equal to the inequality $M_{3322}$ in 
 Ref.~\cite{BGS05} and a violation of it has been observed recently by using 
 photonic qubits \cite{CLBGK15}.
However, while $M_{3322}$ is the only nontrivial facet of the polytope 
 investigated in Ref.~\cite{BGS05}, it is just one of several nontrivial facets 
 in our case.


\subsection{Proof of the inequality in Eq.~\eqref{ineqc}}\label{proof}


Here, we show that for $(n-1)$-ary nonsignaling correlations, the inequality in 
 Eq.~\eqref{ineqc} holds.
We start by letting for some fixed index $0\le \ell < n$,
\begin{subequations}
\begin{align}
 F&=-\sum_b R_{0,b|0,\ell} + \sum_k [ R_{0,0|0,k}-R_{k,0|1,k} ],\\
 X_{1;a|x,y}&=\sum_b(R_{a,b|x,y}-R_{a,b|x,\ell}),\\
 X_{2;b|x,y}&=\sum_a(R_{a,b|x,y}-R_{a,b|0,y}),
\end{align}
\end{subequations}
 where all $R_{a,b|x,y}$ are linearly independent vectors from a real vector 
 space $V$.
Clearly, for any set of correlations, we can find a linear function $\phi\colon 
 V\rightarrow \reals$ with $\phi(R_{a,b|x,y})= P(a,b|x,y)$.
For such a function, $I_b^{(n)}= \phi(F)$ holds and $\phi(X_\tau)= 0$ are all 
 the nonsignaling conditions.
The maximal value of $I_b^{(n)}$ for $(n-1)$-ary nonsignaling correlations is 
 therefore given by
\begin{equation}\label{prim}\begin{split}
 \textstyle\max_{\ell'}
 \max\{ \phi(F) \mid\; & \phi\colon V\rightarrow \reals \text{, linear,}\\
 &\phi(X_\tau) = 0, \text{ for all } \tau, \\
 & \phi(R_{\ell',b|1,y})= 0, \text{ for all } b,y,\\
 & \textstyle\sum_\upsilon \phi(R_\upsilon)= 2n, \text{ and }\\
 & \phi(R_\upsilon)\ge 0, \text{ for all } \upsilon\}.
\end{split}\end{equation}
Since the value of the inner maximization does not depend on the choice of 
 $\ell$, we can choose $\ell=\ell'$.
Equation~\eqref{prim} is a linear program, and the equivalent dual to this 
 program can be written as
\begin{equation}\label{dual}
 \max_\ell
 \min_{t,\boldsymbol\xi, \boldsymbol\eta}
 \set{ t | t\ge \zeta_\upsilon \text{ for all } \upsilon},
\end{equation}
 where $\boldsymbol\zeta$ is the solution of
\begin{equation}
 2 n F - \sum_\tau \xi_\tau X_\tau -\sum_{b,y}\eta_{b,y} R_{\ell,b|1,y}=
 \sum_\upsilon \zeta_\upsilon R_\upsilon.
\end{equation}
To obtain an upper bound in Eq.~\eqref{dual}, we choose $\boldsymbol\eta\equiv 
 2n$ and all $\xi_\tau= 0$, but
$\xi_{1;a|0,k}=4$,
$\xi_{1;k|1,k}=-2n$,
$\xi_{2;b|1,\ell}=-3n+2$, and
$\xi_{2;b|1,k}=-(-1)^bn+2$, for $k\ne \ell$.
This yields $\max_\upsilon \zeta_\upsilon= n-2$ for all $\ell$ and hence the 
 $(n-1)$-ary nonsignaling correlations obey $I_b^{(n)}\le n-2$.


\section{Conclusions}


There was little chance to learn new physics from the recent loophole-free 
 experiments of the Bell inequality \cite{HBD15,GVW15,SMC15,HKB16,W16}.
Years of convincing experiments \cite{FC72,ADR82,WJSWZ98} allowed us to 
 anticipate the conclusions: nature cannot be explained by local realistic 
 theories \cite{Bell64}, there are measurements for which there is not a joint 
 probability distribution \cite{Fine82}, and there are states that are not a 
 convex combination of local states \cite{Werner89}.

Here we have shown how to use Bell-type experiments to gain insights into QT.
In Ref.~\cite{KC16}, it was shown that QT predicts correlations that cannot be 
 explained by nonsignaling correlations produced by fundamentally binary 
 measurements (including Popescu--Rohrlich boxes \cite{PR94}).
We proposed a feasible experiment which will allow us to either exclude all 
 fundamentally binary probabilistic theories or to falsify QT.
If the results of the experiment violate the inequality in Eq.~\eqref{ineqa}, 
 as predicted by QT, then we would learn that no fundamentally binary theory 
 can possibly describe nature.
In addition, it would prove that all involved measurements are genuine 
 three-outcome measurements.
If the inequality in Eq.~\eqref{ineqa} is not violated despite visibilities 
 would \emph{a priori} lead to such a violation, then we would have evidence 
 that QT is wrong at a fundamental level (although being subtle to detect in 
 experiments).
We have also gone beyond Ref.~\cite{KC16} by showing that, for any $n$, already 
 $n$-ary quantum correlations are not fundamentally $(n-1)$-ary.


\begin{acknowledgments}
This work is supported by
Project No.~FIS2014-60843-P, ``Advanced Quantum Information'' (MINECO, Spain), 
with FEDER funds,
the FQXi Large Grant ``The Observer Observed: A Bayesian Route to the 
Reconstruction of Quantum Theory'',
the project ``Photonic Quantum Information'' (Knut and Alice Wallenberg 
Foundation, Sweden),
the Hungarian National Research Fund OTKA (Grants No.~K111734 and No.~KH125096),
the EU (ERC Starting Grant GEDENTQOPT),
and the DFG (Forschungsstipendium KL~2726/2-1).
\end{acknowledgments}



\begin{thebibliography}{99}

\bibitem{FS16}
G. Chiribella and R. W. Spekkens (eds.),
\textit{Quantum Theory: Informational Foundations and Foils},
Fundamental Theories of Physics, Vol.~181
(Springer, Dordrecht, Holland, 2016).

\bibitem{NGHA15}
M. Navascués, Y. Guryanova, M. J. Hoban, and A. Acín,
Almost quantum correlations,
\href{https://doi.org/10.1038/ncomms7288}{Nat. Comm. \textbf{6}, 6288 (2015)}.

\bibitem{KC16}
M. Kleinmann and A. Cabello,
Quantum Correlations Are Stronger Than All Nonsignaling Correlations Produced by $n$-Outcome Measurements,
\href{https://doi.org/10.1103/PhysRevLett.117.150401}{Phys. Rev. Lett. \textbf{117}, 150401 (2016)}.

\bibitem{BHSS13}
P. Busch, T. Heinosaari, J. Schultz, and N. Stevens,
Comparing the degrees of incompatibility inherent in probabilistic physical theories,
\href{https://doi.org/10.1209/0295-5075/103/10002}{EPL \textbf{103}, 10002 (2013)}.

\bibitem{Cabello15}
A. Cabello,
Simple Explanation of the Quantum Limits of Genuine $n$-Body Nonlocality,
\href{https://doi.org/10.1103/PhysRevLett.114.220402}{Phys. Rev. Lett. \textbf{114}, 220402 (2015)}.

\bibitem{HBD15}
B. Hensen, H. Bernien, A. E. Dréau, A. Reiserer, N. Kalb, M. S. Blok, J. Ruitenberg, R. F. L. Vermeulen, R. N. Schouten, C. Abellán, W. Amaya, V. Pruneri, M. W. Mitchell, M. Markham, D. J. Twitchen, D. Elkouss, S. Wehner, T. H. Taminiau, and R. Hanson,
Loophole-free Bell inequality violation using electron spins separated by 1.3 kilometres,
\href{https://doi.org/10.1038/nature15759}{Nature (London) \textbf{526}, 682 (2015)}.

\bibitem{GVW15}
M. Giustina, M. A. M. Versteegh, S. Wengerowsky, J. Handsteiner, A. Hochrainer, K. Phelan, F. Steinlechner, J. Kofler, J.-Å. Larsson, C. Abellán, W. Amaya, V. Pruneri, M. W. Mitchell, J. Beyer, T. Gerrits, A. E. Lita, L. K. Shalm, S. W. Nam, T. Scheidl, R. Ursin, B. Wittmann, and A. Zeilinger,
Significant-Loophole-Free Test of Bell's Theorem with Entangled Photons,
\href{https://doi.org/10.1103/PhysRevLett.115.250401}{Phys. Rev. Lett. \textbf{115}, 250401 (2015)}.

\bibitem{SMC15}
L. K. Shalm, E. Meyer-Scott, B. G. Christensen, P. Bierhorst, M. A. Wayne, M. J. Stevens, T. Gerrits, S. Glancy, D. R. Hamel, M. S. Allman, K. J. Coakley, S. D. Dyer, C. Hodge, A. E. Lita, V. B. Verma, C. Lambrocco, E. Tortorici, A. L. Migdall, Y. Zhang, D. R. Kumor, W. H. Farr, F. Marsili, M. D. Shaw, J. A. Stern, C. Abellán, W. Amaya, V. Pruneri, T. Jennewein, M. W. Mitchell, P. G. Kwiat, J. C. Bienfang, R. P. Mirin, E. Knill, and S. W. Nam,
Strong Loophole-Free Test of Local Realism,
\href{https://doi.org/10.1103/PhysRevLett.115.250402}{Phys. Rev. Lett. \textbf{115}, 250402 (2015)}.

\bibitem{HKB16}
B. Hensen, N. Kalb, M. S. Blok, A. E. Dréau, A. Reiserer, R. F. L. Vermeulen, R. N. Schouten, M. Markham, D. J. Twitchen, K. Goodenough, D. Elkouss, S. Wehner, T. H. Taminiau, and R. Hanson,
Loophole-free Bell test using electron spins in diamond: Second experiment and additional analysis
\href{https://doi.org/10.1038/srep30289}{Sci. Rep. \textbf{6}, 30289 (2016)}.

\bibitem{W16}
W. Rosenfeld, D. Burchardt, R. Garthoff, K. Redeker, N. Ortegel, M. Rau, and H. Weinfurter,
Event-Ready Bell Test Using Entangled Atoms Simultaneously Closing Detection and Locality Loopholes
\href{https://doi.org/10.1103/PhysRevLett.119.010402}{Phys. Rev. Lett. 119, 010402 (2017)}.

\bibitem{PJC15}
H. S. Poh, S. K. Joshi, A. Cerè, A. Cabello, and C. Kurtsiefer,
Approaching Tsirelson's Bound in a Photon Pair Experiment,
\href{https://doi.org/10.1103/PhysRevLett.115.180408}{Phys. Rev. Lett. \textbf{115}, 180408 (2015)}.

\bibitem{CLBGK15}
B. G. Christensen, Y.-C. Liang, N. Brunner, N. Gisin, and P .G. Kwiat,
Exploring the Limits of Quantum Nonlocality with Entangled Photons,
\href{https://doi.org/10.1103/PhysRevX.5.041052}{Phys. Rev. X \textbf{5}, 041052 (2015)}.

\bibitem{VWZ02}
A. Vaziri, G. Weihs, and A. Zeilinger,
Experimental Two-Photon, Three-Dimensional Entanglement for Quantum Communication,
\href{https://doi.org/10.1103/PhysRevLett.89.240401}{Phys. Rev. Lett. \textbf{89}, 240401 (2002)}.

\bibitem{GJVWZ06}
S. Gröblacher, T. Jennewein, A. Vaziri, G. Weihs, and A. Zeilinger,
Experimental quantum cryptography with qutrits,
\href{https://doi.org/10.1088/1367-2630/8/5/075}{New. J. Phys. \textbf{8}, 75 (2006)}.

\bibitem{DLBPA11}
A. C. Dada, J. Leach, G. S. Buller, M. J. Padgett, and E. Andersson,
Experimental high-dimensional two-photon entanglement and violations of generalized Bell inequalities,
\href{https://doi.org/10.1038/nphys1996}{Nat. Phys. \textbf{7}, 677 (2011)}.

\bibitem{Bell64}
J. S. Bell,
On the Einstein Podolsky Rosen paradox,
Physics (Long Island City, NY) \textbf{1}, 195 (1964).

\bibitem{PR94}
S. Popescu and D. Rohrlich,
Quantum nonlocality as an axiom,
\href{https://doi.org/10.1007/BF02058098}{Found. Phys. \textbf{24}, 379 (1994)}.

\bibitem{NPA07}
M. Navascués, S. Pironio, and A. Acín,
Bounding the Set of Quantum Correlations,
\href{https://doi.org/10.1103/PhysRevLett.98.010401}{Phys. Rev. Lett. \textbf{98}, 010401 (2007)}.

\bibitem{porta}
POlyhedron Representation Transformation Algorithm,
retrieved from \url{http://porta.zib.de/}.

\bibitem{ppl}
Parma Polyhedra Library,
retrieved from \url{http://bugseng.com/products/ppl/}.

\bibitem{BGS05}
N. Brunner, N. Gisin, and V. Scarani,
Entanglement and non-locality are different resources,
\href{https://doi.org/10.1088/1367-2630/7/1/088}{New. J. Phys. \textbf{7}, 88 (2005)}.

\bibitem{FC72}
S. J. Freedman and J. F. Clauser,
Experimental Test of Local Hidden-Variable Theories,
\href{https://doi.org/10.1103/PhysRevLett.28.938}{Phys. Rev. Lett. \textbf{28}, 938 (1972)}.

\bibitem{ADR82}
A. Aspect, J. Dalibard, and G. Roger,
Experimental Test of Bell's Inequalities Using Time-Varying Analyzers,
\href{https://doi.org/10.1103/PhysRevLett.49.1804}{Phys. Rev. Lett. \textbf{49}, 1804 (1982)}.

\bibitem{WJSWZ98}
G. Weihs, T. Jennewein, C. Simon, H. Weinfurter, and A. Zeilinger,
Violation of Bell's Inequality under Strict Einstein Locality Conditions,
\href{https://doi.org/10.1103/PhysRevLett.81.5039}{Phys. Rev. Lett. \textbf{81}, 5039 (1998)}.

\bibitem{Fine82}
A. Fine,
Hidden Variables, Joint Probability, and the Bell Inequalities,
\href{https://doi.org/10.1103/PhysRevLett.48.291}{Phys. Rev. Lett. \textbf{48}, 291 (1982)}.

\bibitem{Werner89}
R. F. Werner,
Quantum states with Einstein-Podolsky-Rosen correlations admitting a hidden-variable model,
\href{https://doi.org/10.1103/PhysRevA.40.4277}{Phys. Rev. A \textbf{40}, 4277 (1989)}.

\end{thebibliography}
\end{document}